\documentclass[linenumbers,superscriptaddress,twocolumn,prc,showpacs]{revtex4-1}

\usepackage{amsmath}
\usepackage{amssymb}
\usepackage{datetime}
\usepackage{graphicx}
\usepackage[latin1]{inputenc}
\usepackage{hyperref}
\usepackage{dcolumn}
\usepackage{units}
\usepackage{upgreek}
\usepackage{url}

\hypersetup{%
  pdfstartview=FitH,%
  bookmarksopen=true,%
  breaklinks=true,%
  colorlinks=true,%
  linkcolor=blue,anchorcolor=blue,%
  citecolor=blue,filecolor=blue,%
  menucolor=blue,%
  urlcolor=blue%
}

\begin{document}

\newcommand{\tsup}[1]{\textsuperscript{#1}}
\newcommand{\tcol}[1]{\multicolumn{1}{c}{#1}}


\renewcommand{\doi}[1]{\href{http://dx.doi.org/#1}{$\Rsh$}}
\newcommand{\urlref}[1]{\href{#1}{$\Rsh$}}



\title{Nuclear Level Density and Gamma-Ray Strength Function of \tsup{43}Sc}

\author{A.~B\"urger}
\affiliation{University of Oslo, Department of Physics, N-0316 Oslo, Norway}
\author{A.C.~Larsen}
\email[Email address: ]{a.c.larsen@fys.uio.no}
\affiliation{University of Oslo, Department of Physics, N-0316 Oslo, Norway}
\author{S.~Hilaire}
\affiliation{CEA, DAM, DIF, F-91297 Arpajon, France}
\author{M.~Guttormsen}
\affiliation{University of Oslo, Department of Physics, N-0316 Oslo, Norway}
\author{S.~Harissopulos}
\affiliation{Institute of Nuclear Physics, NCSR ``Demokritos'', 153.10 Aghia Paraskevi, Athens, Greece}
\author{M.~Kmiecik}
\affiliation{Institute of Nuclear Physics PAN, Krak{\'o}w, Poland}
\author{T.~Konstantinopoulos}
\affiliation{Institute of Nuclear Physics, NCSR ``Demokritos'', 153.10 Aghia Paraskevi, Athens, Greece}
\author{M.~Krti\v{c}ka}
\affiliation{Institute of Nuclear and Particle Physics, Charles University, Prague, Czech Republic}
\author{A.~Lagoyannis}
\affiliation{Institute of Nuclear Physics, NCSR ``Demokritos'', 153.10 Aghia Paraskevi, Athens, Greece}
\author{T.~L\"onnroth}
\affiliation{Physics, Department of Natural Sciences, \AA{}bo Akademi University, FIN-20500 \AA{}bo, Finland}
\author{K.~Mazurek}
\affiliation{Institute of Nuclear Physics PAN, Krak{\'o}w, Poland}
\author{M.~Norrby}
\affiliation{Physics, Department of Natural Sciences, \AA{}bo Akademi University, FIN-20500 \AA{}bo, Finland}
\author{H.T.~Nyhus}
\affiliation{University of Oslo, Department of Physics, N-0316 Oslo, Norway}
\author{G.~Perdikakis}
\altaffiliation[Present address: ]{National Superconducting Cyclotron Laboratory, Michigan State University, East Lansing, Michigan 48824, USA}
\affiliation{Institute of Nuclear Physics, NCSR ``Demokritos'', 153.10 Aghia Paraskevi, Athens, Greece}
\author{S.~Siem}
\affiliation{University of Oslo, Department of Physics, N-0316 Oslo, Norway}
\author{A.~Spyrou}
\altaffiliation[Present address: ]{National Superconducting Cyclotron Laboratory, Michigan State University, East Lansing, Michigan 48824, USA}
\affiliation{Institute of Nuclear Physics, NCSR ``Demokritos'', 153.10 Aghia Paraskevi, Athens, Greece}
\author{N.U.H.~Syed}
\affiliation{University of Oslo, Department of Physics, N-0316 Oslo, Norway}


 \begin{abstract}%
   The nuclear level density and the $\gamma$-ray strength function
   have been determined for \tsup{43}Sc in the energy range up to
   $\unit[2]{MeV}$ below the neutron separation energy using the Oslo
   method with the \tsup{46}Ti$(p,\alpha)$\tsup{43}Sc reaction.
   A comparison to \tsup{45}Sc shows that the level density of
   \tsup{43}Sc is smaller by an approximately constant factor of two.
   This behaviour is well reproduced in a microscopical/combinatorial
   model calculation.
   The $\gamma$-ray strength function is showing an increase at low
   $\gamma$-ray energies, a feature which has been observed in several
   nuclei but which still awaits theoretical explanation.
\end{abstract}



\maketitle



\section{Introduction}
\label{sec:intro}

Network calculations aiming to reproduce isotopic abundances observed
in stars, or predictions of isotope productions in nuclear power
plants require good knowledge of nuclear level densities and
$\gamma$-ray transition rates for many nuclei and over a large range
of excitation energies to calculate the relevant cross sections.
Up to a certain excitation energy, it is feasible to perform
spectroscopic measurements on all individual nuclear excited states
and to determine at least some of their properties.
But at higher excitation energies, the spacing between nuclear levels
may become very small, which does not allow to resolve all individual
levels.
A continuing effort has since long been devoted both in experiment and
theory to the study of level densities and $\gamma$-ray strengths
function also in this region of quasi-continuum.
Despite these efforts, the amount of available experimental data is
relatively small.
Therefore, network calculations often have to rely on models to
compensate for the lack of measured values, and models are difficult
to validate without experimental data to compare with.

The nuclear physics group at the University of Oslo has performed many
experiments using the Oslo method to determine nuclear level densities
and $\gamma$-ray strength functions of many isotopes throughout the
nuclear
chart~\cite{Chankova06,Lar07,Algin08,Agvaanluvsan09,Syed09a,Nyhus10}.
In the present work, the Oslo method has been used for the first time
on a nucleus produced in a $(p,\alpha)$ reaction to determine the
level density and the $\gamma$-ray strength function of \tsup{43}Sc.
Previously published data for \tsup{45}Sc~\cite{Lar07}, produced in
the (\tsup{3}He,\tsup{3}He$^\prime$) reaction, allow the comparison of
two relatively light isotopes with $\Delta A=2$.

In the next section, the experimental setup and the analysis procedure
are described, followed by a discussion of the experimental results in
sections~\ref{sec:nld} and \ref{sec:strength}.
In section~\ref{sec:summary}, we conclude with a summary.


\section{Experiment and Data Analysis}
\label{sec:setup}

The experiment was performed at the cyclotron laboratory of the
University of Oslo.
A proton beam with an energy of $\unit[32]{MeV}$ impinged on a Ti
target of $\unit[3]{mg/cm^2}$ thickness with an enrichment of
$\unit[86]{\%}$ \tsup{46}Ti.
The main impurities were \tsup{48}Ti ($\unit[10.6]{\%}$), \tsup{47}Ti
($\unit[1.6]{\%}$), \tsup{50}Ti ($\unit[1.0]{\%}$), and \tsup{49}Ti
($\unit[0.8]{\%}$).
Eight silicon $\Delta E-E$ particle telescopes with a total geometric
efficiency of about $\unit[1.3]{\%}$ were placed in forward direction
at $\unit[5]{cm}$ distance behind the target at an angle of
$\unit[45]{^\circ}$ with respect to the beam axis.
The target was surrounded by the $\gamma$-ray detector array CACTUS,
consisting of 28 collimated NaI(Tl) scintillator crystals covering
about $\unit[15]{\%}$ of $4\pi$.

Using the specific energy losses in the thin ($\unit[140]{\upmu m}$)
$\Delta E$ and the thick ($\unit[1500]{\upmu m}$) $E$ particle
detectors, $\alpha$ ejectiles were identified to select the
\tsup{46}Ti$(p,\alpha)$\tsup{43}Sc reaction channel.
From the known $Q$-values, the reaction kinematics and the energy
losses in the materials passed by the $\alpha$ particles, the initial
excitation energies $E_i$ of the \tsup{43}Sc nuclei could be
reconstructed with an accuracy of about $\unit[700]{keV}$ FWHM.

The difference in total energy deposit in the Si detectors between
\tsup{43}Sc (produced from the main target component, \tsup{46}Ti) and
\tsup{45}Sc (produced from the main impurity, \tsup{48}Ti) in the
respective $(p,\alpha)$ reactions is only about $\unit[0.5]{MeV}$ for
the ground states.
With the present experimental setup, it is not possible to separate
the reactions on the two target components and a certain level of
background from \tsup{45}Sc cannot be removed from the spectra for
\tsup{43}Sc. 

From the amount of impurities in the target, one would expect 
that the contribution from these impurities should not exceed 
$\approx 14\%$. This assumption is supported by $(p,t)$ data from 
the same experiment (see Fig.~2 in Ref.~\cite{Lar2012}). Here, it is
clear that the main impurity is stemming from $(p,t)$\tsup{46}Ti, which
is of the order of $10\%$. In addition, from calculations of 
differential cross sections at $\unit[45]{^\circ}$ for the 
$(p,\alpha)$\tsup{43,45}Sc reactions, and from the cross-section data of
Ref.~\cite{AbouZeid1980}, we find no significant difference
in neither the absolute value nor the shape of the estimated $\alpha$
spectra. Therefore, it seems reasonable to believe that the background
from \tsup{45}Sc in the present data is of the same order as the 
amount of \tsup{48}Ti in the target. 

\begin{figure}[tb]
  \centering
  \includegraphics[width=\linewidth]{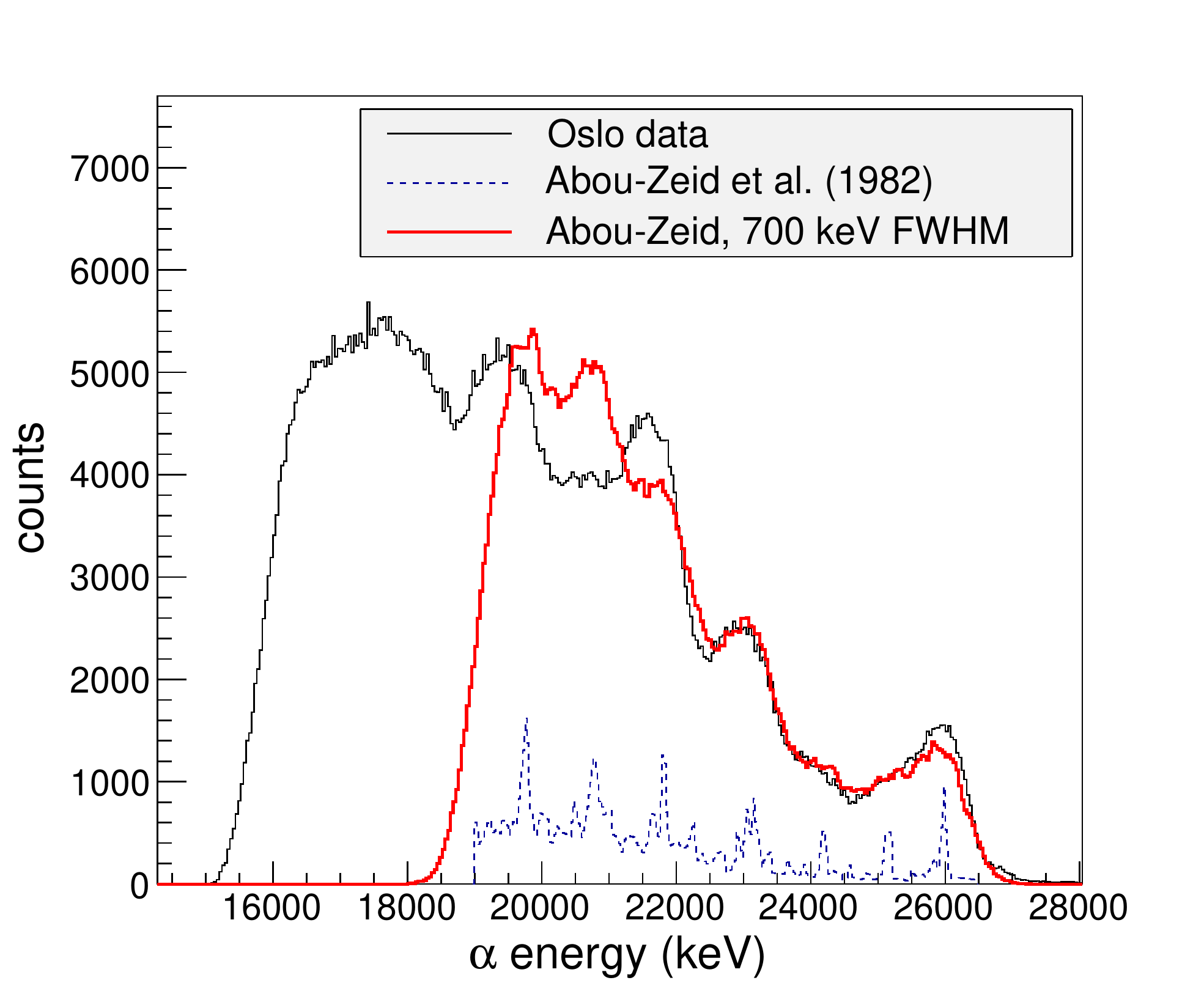}
  \caption{(color online) 
    Alpha spectra from the  \tsup{46}Ti($p,\alpha$)\tsup{43}Sc reaction:
    Oslo data from one particle telescope (black solid line), 
    data from Abou-Zeid \textit{et al.} scaled a factor of 10 
    (blue, dashed line, from 
    Ref.~\cite{AbouZeid1980}), and the Abou-Zeid data folded with
    a Gaussian of 700 keV FWHM, and scaled by a factor of 40 (red line).}
  \label{fig:alphaspectra}
\end{figure}

A comparison with the \tsup{43}Sc $\alpha$-particle spectrum from
Ref.~\cite{AbouZeid1980}, folded with the present detector resolution,
shows very good agreement for $\alpha$ energies above $\approx 22$ MeV, 
see Fig.~\ref{fig:alphaspectra}.
This further indicates that the contribution from \tsup{45}Sc is rather small.

We observe deviations between our data and the Abou-Zeid data for 
$E_{\alpha} < 21$ MeV. In particular, this is so for the peaks centered at 
$\approx 21.6$ MeV and $\approx 19.5$ MeV in our data, and the peak at 
$\approx 20.7$ MeV in the Abou-Zeid data. The former ones are coming by 
the $^{16}$O($p,\alpha$)$^{13}$N reaction, as the Ti target had a 
layer of TiO$_{2}$ on the surface. However, there is no obvious reason 
for the difference of the latter peak. Possibly, the different beam
energy and scattering angle (our data cover angles between 
$\approx 43-47^{\circ}$) could account for the observed deviation.

As a consequence of the \tsup{45}Sc contribution, 
some smoothing effects on the extracted quantities are expected. 

The excited \tsup{43}Sc nuclei will emit cascades of $\gamma$ rays to
decay to their ground states.
The spectra of these $\gamma$-ray energies $E_\gamma$ were measured in
coincidence with the $\alpha$ particles, and a matrix $E_i$
vs. $E_{\gamma}$ was constructed after correcting for the NaI response
function as described in \cite{Sch00}.
This matrix of unfolded $\gamma$-ray spectra was normalized such that
for each initial excitation energy $E_i$, the integral over all
$\gamma$-ray energies measured in coincidence with this excitation
energy equaled the average $\gamma$-ray multiplicity observed in this
excitation energy bin.
The average multiplicity was determined as $\langle M\rangle =
E_i/\langle E_\gamma \rangle$ with the average $\gamma$-ray energy
$\langle E_\gamma \rangle$ for the excitation energy bin
$E_i$~\cite{Rekstad1983}.
The first-generation method \cite{Gut87} was then applied on this
matrix to extract a matrix $P$ containing the spectrum of primary
$\gamma$-ray energies for each initial excitation energy bin $E_i$.
A fundamental assumption for the first-generation method is that the
$\gamma$-ray spectrum emitted from each excitation energy bin is
independent of how the states in this bin were populated -- by
$\gamma$ decay from higher excited states or by population in the
$(p,\alpha)$ reaction.

From the matrix $P$, both the shape of the level density $\rho(E_f)$
and the shape of the $\gamma$-ray strength function $f(E_\gamma)$ can
be extracted as described in \cite{Sch00}.
As explained there, this extraction can only be performed if the
$\gamma$-ray strength function only depends on the $\gamma$-ray
energy, but not on the excitation energy (the generalized Brink-Axel
hypothesis \cite{Brink1955,Axel1962}), and the transition probability
from an initial state $i$ to a final state $f$ (with excitation
energies $E_i$ and $E_f$, respectively) can be factorized into the
level density at the final state, $\rho(E_f)$, and the $\gamma$
transmission coefficient, ${\cal T}(E_i-E_f)$.
Furthermore it is assumed in the following that dipole radiation is
predominant and that one can write ${\cal T}(E_\gamma) = 2\pi
f(E_\gamma) E_\gamma^3$.
The results obtained at this point are only the functional forms of
$\rho$ and $f$ in the sense that the matrix $P$ can be equally fitted
to other pairs of $\rho'$ and $f'$ obtained by the transformations:
\begin{linenomath}
  \begin{align}
    \label{eq:tansformations}
    \rho'(E_f)  &= A \exp(\alpha E_f) \rho(E_f) \\
    f'(E_\gamma) &= B \exp(\alpha E_\gamma) f(E_\gamma)
  \end{align}
\end{linenomath}
for any positive values of $A$, $B$, and any value of
$\alpha$~\cite{Sch00}.

To determine appropriate values for these coefficients, the level
density and the $\gamma$-ray strength function must be normalized
using data from other sources.
The parameters $A$ and $\alpha$ were determined using two level
density values:
One of them is the counted level density from discrete-line
spectroscopy at low excitation energy, where it has been assumed that
all levels have been observed (green region in Fig.~\ref{fig:nld}).
The second one is the level density derived from resonance spacings at
average energy $E_n$, slightly above a particle separation energy.
This value is extrapolated to lower excitation energies using a scaled
back-shifted Fermi gas (BSFG) model~\cite{EgiBuc05} to bridge the gap
between the maximum energy for which $\rho(E_f)$ can be determined in
the experiment and $E_n$ (red line and region in Fig.~\ref{fig:nld}).
While no neutron resonance data are available for \tsup{43}Sc, some
information on proton resonances is tabulated in~\cite{Sukh0x}.
To perform the normalization, it has been assumed that the tentative
spin assignments in~\cite{Sukh0x} are correct, and that the
distribution of unknown spin-parity values equals the distribution of
known spin-parity values.
The normalization point for the level density has then been obtained
by counting the levels in the excitation energy region around
$E_n=\unit[7]{MeV}$.
The BSFG parameters for the extrapolation are the same as used for
\tsup{45}Sc in~\cite{Lar07}: the level density parameter was
$a=\unit[4.94]{MeV^{-1}}$, the back-shift parameter
$E_1=\unit[-2.55]{MeV}$.
In addition, the curve was scaled with a factor $\eta=0.585$ to match
the level density normalization point for \tsup{43}Sc as obtained from
the proton resonance data.
This particular BSFG parametrization was chosen to allow a comparison
with the data for \tsup{45}Sc from Ref.~\cite{Lar07}.

A third normalization point is necessary to fix the parameter $B$ for
the $\gamma$-ray strength function scale.
If available, data on the average total radiative width could be used
for this purpose~\cite{Voi01,Gut05}.
Such data are, however, not available for \tsup{43}Sc.
Therefore, estimated $\gamma$-ray strength function values for
\tsup{46}Sc have been used in exactly the same way as for \tsup{45}Sc
in \cite{Lar07}:
the normalization value is the sum of the $E1$ and $M1$ strength
function values for \tsup{46}Sc from Ref.~\cite{KopeckyUhl1995}.
The use of the \tsup{46}Sc value is justified if it is assumed that
the $\gamma$-ray strength functions for \tsup{43}Sc and \tsup{46}Sc
(and \tsup{45}Sc) are not very different in scale.


\section{Nuclear Level Density}
\label{sec:nld}

\begin{figure}[tb]
  \centering
  \includegraphics[width=\linewidth]{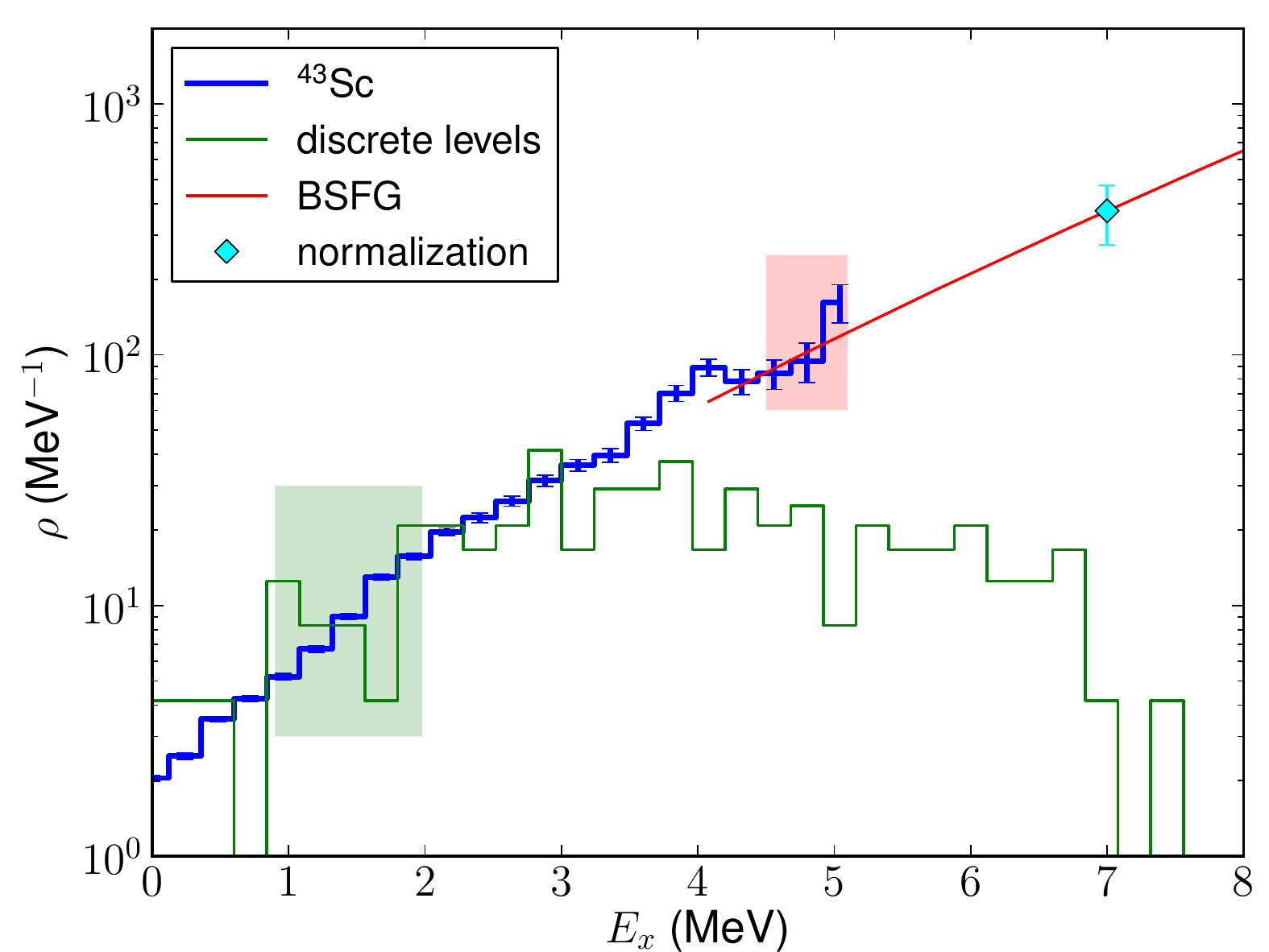}
  \caption{(color online) 
    Experimental level density for \tsup{43}Sc.
    The experimental curve (blue steps) is normalized to discrete
    levels (green steps, fitted in green region) and to proton
    resonance spacings (cyan diamond) extrapolated using a BSFG model
    (red line, fitted in the red region).}
  \label{fig:nld}
\end{figure}
Figure~\ref{fig:nld} shows the level density curve obtained for
\tsup{43}Sc after the normalization as described in the previous
section.
The level density normalization point at $E_n=\unit[7]{MeV}$ is
$\rho(E_n)=\unit[375]{MeV^{-1}}$ with an estimated uncertainty of
$\Delta \rho(E_n)=\unit[100]{MeV^{-1}}$.
The uncertainties for the experimental data points in this figure are
estimated mainly based on the number of counts in the $E_i$
vs. $E_\gamma$ matrices (see~\cite{Sch00}).
They do, in particular, not include the uncertainty from the
normalization.

\begin{figure}[tb]
  \centering
  \includegraphics[width=\linewidth]{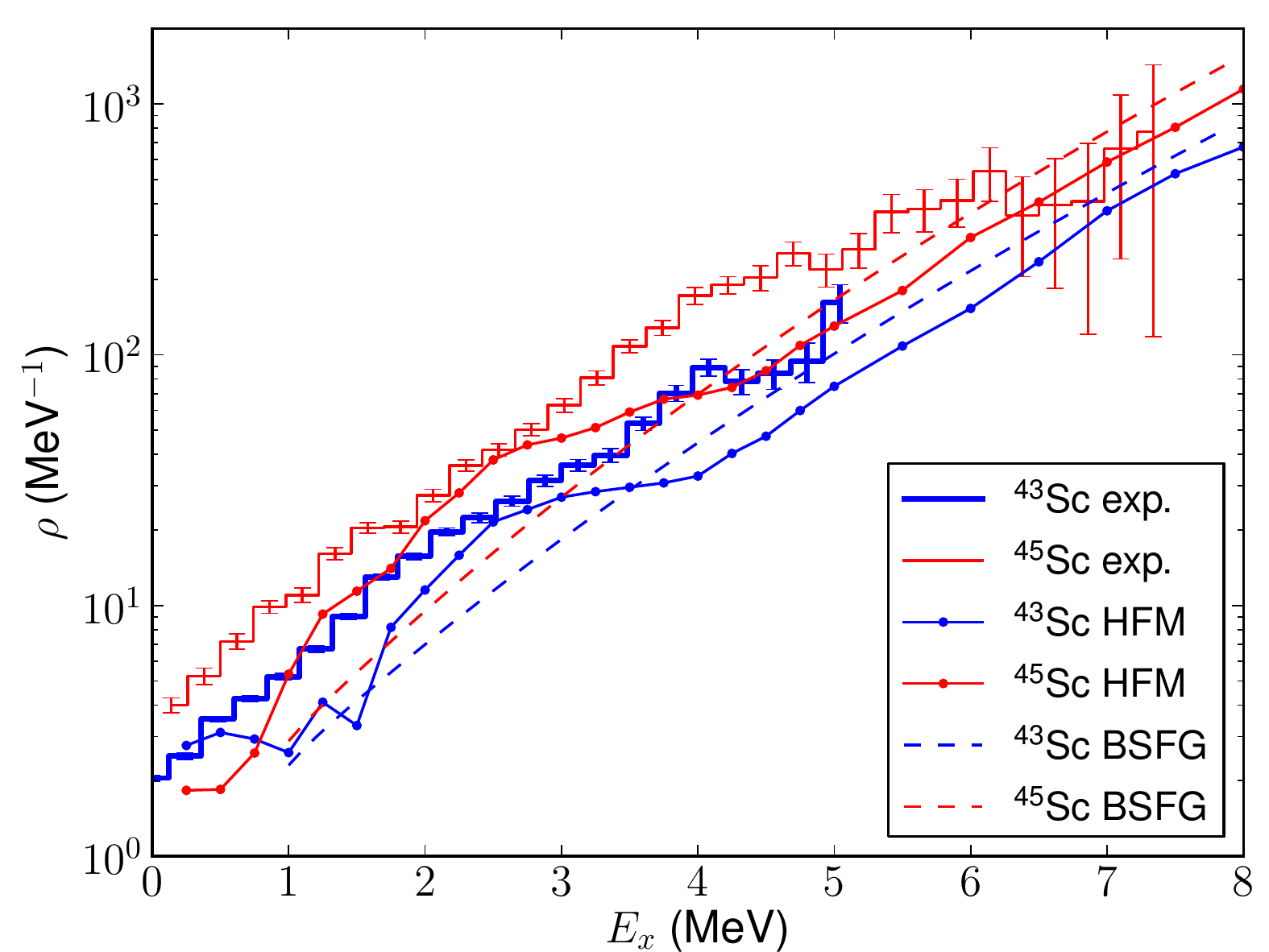}
  \caption{(color online) 
    Level density comparison.
    The experimental level density for \tsup{43}Sc (blue steps) is
    compared to the experimental level density of \tsup{45}Sc (red
    steps) \cite{Lar07} and to combinatorial model calculations for
    these two nuclei (blue and red lines for \tsup{43}Sc and
    \tsup{45}Sc, respectively) see text. In addition, BSFG
    calculations are included \tsup{43}Sc and \tsup{45}Sc(blue and red
    dashed lines, respectively).}
  \label{fig:nld_compare}
\end{figure}
In Fig.~\ref{fig:nld_compare}, the experimental level density is
compared to the previously published level density curve for
\tsup{45}Sc \cite{Lar07}.
It appears that, in logarithmic scale, the two level density curves
are more or less parallel to each other: the level density for
\tsup{45}Sc is larger by a factor of about 2 for a large excitation
energy range.
A similar behavior has been found in heavier nuclei:
near closed shells, the level density of nuclei of the same element
with masses $A$ and $(A+2)$ differs
significantly~\cite{Siem2009,Chankova06,Syed09a}.
For mid-shell nuclei, on the other hand, the level density of $\Delta
A=2$ neighbors of the same element is very similar or almost identical
in scale~\cite{Agvaanluvsan04,Guttormsen03}.



It is not obvious where this increase by a factor 2 between $A=43$ and
$A=45$ originates from.
In case of spherical nuclei with pronounced $N = 20$ and $N = 28$
shell gaps, only a few active particles in the $f_{7/2}$ orbitals
would be responsible for the number of levels, namely 1~$\pi$ and
2~$\nu$ for \tsup{43}Sc, and 1~$\pi$ and 4~$\nu$ for \tsup{45}Sc.
In this picture, one could expect many more configurations at one and
the same excitation energy for \tsup{45}Sc compared to \tsup{43}Sc.
However, both isotopes have an $I^\pi=3/2^+$ state just above the
$7/2^-$ ground state, indicating that the $d_{3/2}$ hole orbital is
close to the $f_{7/2}$, which can be explained by a quadrupole
deformation of $\epsilon_2 \approx 0.23$ as shown in the Nilsson
single particle scheme of Ref.~\cite{Lar07}.
These calculations show a rather uniform distribution of $\Omega^\pi$
Nilsson orbitals, and one could expect very similar level densities
for \tsup{45}Sc and \tsup{43}Sc.
On the other hand, it is well established that \tsup{45}Sc exhibits
coexistence of prolate and weakly oblate (nearly spherical) rotational
bands~\cite{Bednarczyk1997}.
Since the level density includes all types of configurations with
various spins and parities, one has to expect contributions from both
shapes, where the near-spherical shape might drive towards a large
level density ratio and the deformed shape towards a small level
density ratio between \tsup{43}Sc and \tsup{45}Sc.
The situation is complex and it is difficult to present simple
arguments to explain the experimentally observed level density ratio
of $\approx 2$.


Figure~\ref{fig:nld_compare} includes calculations of level densities
for \tsup{43}Sc and \tsup{45}Sc using the phenomenological BSFG model.
For these curves, the global parametrization from \cite{ko08} was
used (which is different from the parameters used for the
normalization in Sec.~\ref{sec:setup} and Fig.~\ref{fig:nld}).
This parametrization includes shell effects via nuclear masses, which
enter the calculation of the level density parameter $a$.
The resulting ratio of level densities is $1.5$ at
$E_x=\unit[3]{MeV}$, slightly smaller than the ratio of $2$ seen in
experiment.
Generally, the two BSFG calculations tend to underestimate the level
density below $E_x=\unit[6]{MeV}$.

Figure~\ref{fig:nld_compare} also includes theoretical level density
curves derived from calculations using the combinatorial HFM model
described in Ref.~\cite{gorhil08}.
These theoretical level densities were retrieved from
Ref.~\cite{RIPL3}.
As explained in Ref.~\cite{gorhil08}, a meaningful comparison of the
theoretical predictions $\rho_{\rm HFM}$ with the experimental data
$\rho_{\rm exp}$ requires a normalization of $\rho_{\rm HFM}$ to the
level density value used to normalize the experimental level densities
at a given energy $E_{n}$.
Following the normalization recipe of Ref.~\cite{ko08}, we thus
determine for both of \tsup{43}Sc and \tsup{45}Sc a normalization
parameter $c$ such that
\begin{linenomath}
  \begin{equation}
    \label{eq:norm_theo}
    \rho_{\rm HFM}(E_n) \times \exp(c \sqrt{E_n})=\rho_{\rm exp}(E_n)
    \mbox{,}
  \end{equation}
\end{linenomath}
\noindent and then plot in Fig.~\ref{fig:nld_compare} the normalized
values, i.e.
\begin{linenomath}
  \begin{equation}
    \rho_{\rm HFM}(E_x) \times \exp(c \sqrt{E_x})
  \end{equation}
\end{linenomath}
as a function of $E_x$.
\begin{table}
  \centering
  \begin{ruledtabular}
    \begin{tabular}{ldddd}
      \tcol{Nucleus} & \tcol{$E_n$} & \tcol{$\rho(E_n)$} & \tcol{$\rho_{\rm HFM}(E_n)$} & \tcol{$c$} \\
      & \tcol{(MeV)} & \tcol{(MeV$^{-1}$)} & \tcol{(MeV$^{-1}$)} & \\
      \hline
      \tsup{43}Sc & 7.0   &  375 & 1022  & -0.379 \\
      \tsup{45}Sc & 9.904 & 3701 & 11470 & -0.359 \\
    \end{tabular}
  \end{ruledtabular}
  \caption{Microscopical model normalization parameters. The values for the level density
    normalization energy, $E_n$, and density value, $\rho(E_n)$, for \tsup{45}Sc are from
    Ref.~\cite{Lar07}, and the values for $\rho_{\rm HFM}$ were interpolated
    from \cite{RIPL3}.}
  \label{tab:norm_hfm}
\end{table}
In Fig.~\ref{fig:nld_compare}, we chose zero pairing shift and
obtained values for $c$ from eq.~\eqref{eq:norm_theo} as listed in
Table~\ref{tab:norm_hfm}.
The normalized HFM curves nearly reproduce the parallel trend of the
level density curves and the ratio between them with a significant
increase of the \tsup{45}Sc level densities with respect to that of
\tsup{43}Sc, but they underestimate the level densities for both
nuclei.
The main qualitative differences between the HFM calculation and
experimental data are at excitation energies below $\unit[1.5]{MeV}$,
where the calculation does neither reproduce the level densities as
obtained from discrete level counting nor their ratio, and in the
excitation energy range between around $1.5$ and $\unit[4]{MeV}$ where
the model predicts a local increase in the level density for both
nuclei which is not seen in experiment.

At excitation energies below $\unit[2]{MeV}$, the HFM curves show more
structure than the experimental curves.
One possible explanation is the experimental energy resolution.
Another possibility to explain this mismatch is the too approximate
treatment of the coupling between particle-hole and vibrational
excitations implemented in the combinatorial HFM model.
To check this hypothesis, we tested a simplistic model to mimic a more
realistic particle-vibration coupling resulting in a spreading of the
coupled states by an arbitrarily chosen energy of the order of a few
hundred keV.
The HFM curves obtained using such a simplistic treatment show, as
expected, less structure and better agreement with the shape of the
experimental data.
The tested modifications are, however, completely arbitrary and have
to be investigated and refined in future work before including them in
the general HFM calculations.


\section{Gamma-Ray Strength Function}
\label{sec:strength}

\begin{figure}[tb]
  \centering
  \includegraphics[width=\linewidth]{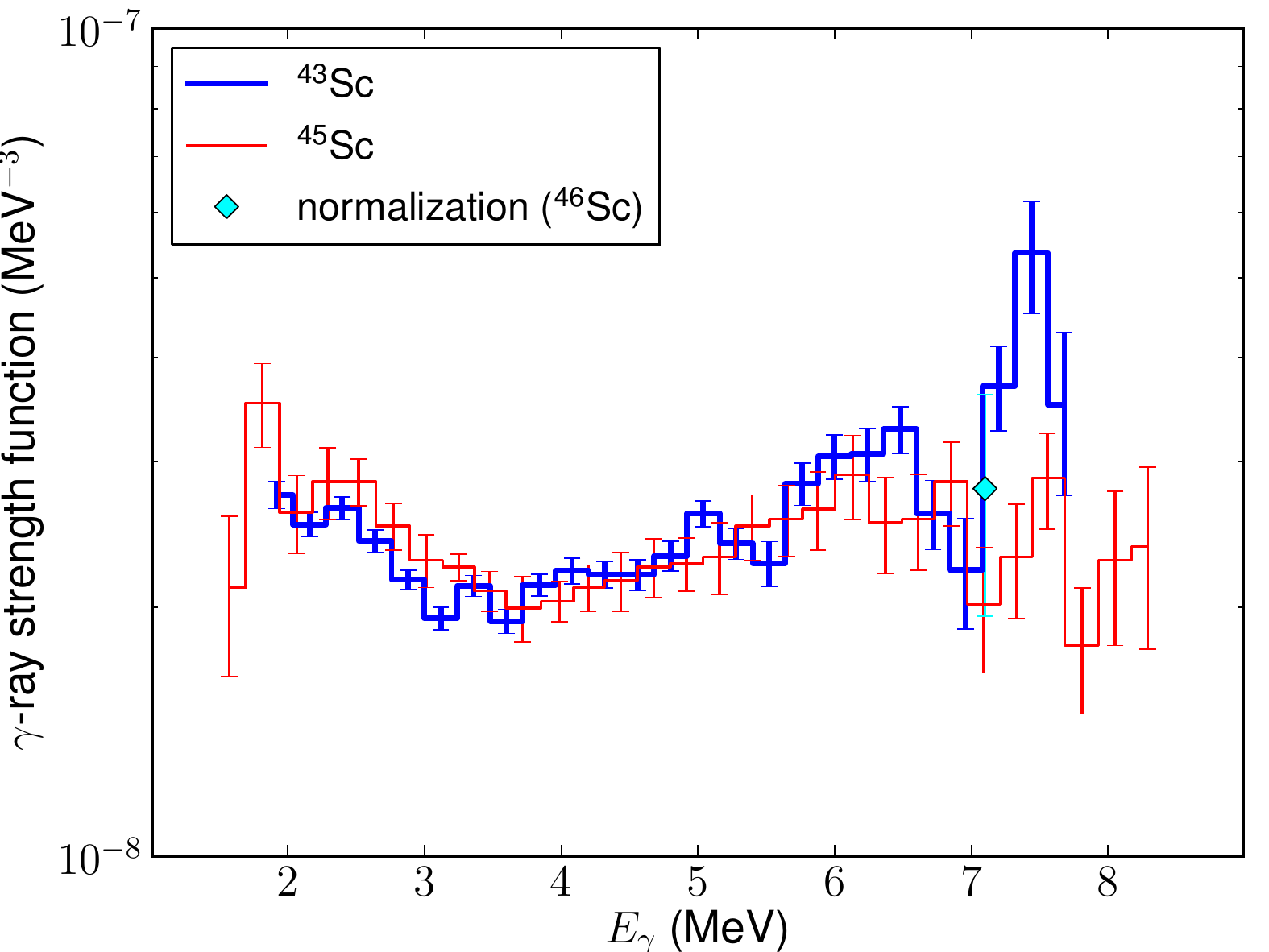}
  \caption{(color online) 
    Gamma-ray strength function for \tsup{43}Sc.
    The experimental curve (blue steps) is shown together with the
    $\gamma$-ray strength function for \tsup{45}Sc (red steps).
    The normalization data point from \tsup{46}Sc is also shown
    (cyan diamond).
  }
  \label{fig:strength}
\end{figure}
Figure~\ref{fig:strength} shows the experimental curves of the
$\gamma$-ray strength function for \tsup{43}Sc, together with the
experimental data for \tsup{45}Sc~\cite{Lar07}.
As for the level density, the uncertainties for the experimental data
points are estimated mainly based on the number of counts in the $E_i$
vs. $E_\gamma$ matrices.
The similarity of the shapes of the measured $\gamma$-ray strength
functions of the $\Delta A=2$ neighbors is astonishing.
A common feature of the curves is that they both show a minimum at
around $\unit[3.5]{MeV}$ and an increase of the $\gamma$-ray strength
function for lower $\gamma$-ray energies.
Similar behavior has been observed in other nuclei and using
different experimental approaches \cite{Syed09b,
  Voinov2004,Gut05,Voinov2010}.

A possible explanation for the case of light nuclei is the typically
low level density at low excitation energy, in particular the scarcity
of higher-spin states, and the dominance of $E1$ radiation.
For a higher-spin state -- which can be populated in the
particle-induced reaction --, the de-excitation then needs multiple,
smaller-energy steps to reach one of the available low-spin states at
low excitation energy~\cite{Larsen2011}.

Phenomenological models describing such $\gamma$-ray strength
functions shows that the increased $\gamma$-ray strength for low
$E_\gamma$ may have important effects on radiative neutron capture
cross sections and thus on r-process nucleosynthesis
calculations~\cite{LarsenGoriely2010}.


\section{Summary}
\label{sec:summary}

The nuclear level density and the $\gamma$-ray strength function of
\tsup{43}Sc have been determined experimentally using the Oslo method.
There is an almost constant factor between the level densities of
\tsup{43}Sc and \tsup{45}Sc, a behavior similar to what has been
observed in heavier nuclei in the vicinity of shell closures.
The parallel evolution of the level densities of the two $\Delta A=2$
isotope neighbors can be nearly reproduced within a combinatorial
model for a large excitation energy range.
The $\gamma$-ray strength function for \tsup{43}Sc is surprisingly
similar to the one of \tsup{45}Sc, and it shows an increase at low
$\gamma$-ray energy which cannot be explained theoretically as of yet.


\appendix

\section*{Acknowledgments}
\label{apx:ack}

The authors wish to thank E.~A.~Olsen and J.~Wikne for excellent
experimental conditions.
Financial support from the Research Council of Norway (Norges
forsk\-nings\-r{\aa}d, project 180663) is gratefully acknowledged.

\bibliographystyle{prsty}
\bibliography{sc43}

\end{document}